\documentclass[amsmath, amssymb, preprint, nofootinbib]{revtex4-1}
\usepackage{graphicx}
\usepackage{units}
\usepackage{subfig}

\begin{document}

\newcommand{\ds}{^{}}
\newcommand{\Etot}{E_\text{tot}}
\newcommand{\bra}[1]{\langle #1 |}
\newcommand{\ket}[1]{| #1 \rangle}
\newcommand{\braket}[2]{\langle #1 | #2 \rangle}
\newcommand{\ident}{\openone}
\newcommand{\diff}[1]{{\rm d}#1}
\newcommand{\Ex}[1]{\langle#1\rangle}
\newcommand{\bigO}[1]{\mathcal{O}(#1)}

\title{Spherical harmonics for multiparticle final states}

\author{Keith Pedersen}
\email{kpeders1@hawk.iit.edu}
\author{Zack Sullivan}
\email{Zack.Sullivan@iit.edu}
\affiliation{Illinois Institute of Technology, Chicago, IL, USA}

\begin{abstract}
We examine collider physics via the spectral power $H_l\ds$,
which utilizes much more detector information than sequential jet reconstruction.
We use $H_l\ds$ to define a novel jet definition and use it to 
reconstruct both $e^+e^-\to q\bar{q}g$ and $e^+e^-\to t\bar{t}$ final states.
We find that a significant amount of pileup can be trivially subtracted from $H_l\ds$ jets,
provided that the ``shape'' of pileup can be determined ahead of time.
Finally, we find that future progress with $H_l\ds$ jets will require 
the addition of degrees of freedom accounting for jet ``shape'' (e.g.\ radial energy distribution).
This should allow the measurement of jet shapes using 
much more information than traditional techniques.
\end{abstract}

\maketitle

\section*{Introduction}

For the  high-luminosity (HL) upgrade to CERN's Large Hadron Collider (LHC) to 
be a success, jet reconstruction and jet observables must be robust to intense pileup.
Pileup is already problematic at the LHC, and while many solutions have been developed, 
they generally augment sequential jet reconstruction algorithms like anti-$kt$.
This fundamentally limits robust jet physics to a reductionist, bottom-up approach
--- survey every 2-particle correlation, but determine the next move using only one ``special'' correlation.
The complementary approach, a holistic, top-down scheme which 
simultaneously uses all correlations in an event 
(e.g.\ event shape variables like thrust and sphericity),
traditionally project the event into one (or a handful) of scalars
--- not incredibly useful for extracting event kinematics.

One can imagine a middle ground where much more information is used at 
every stage in the algorithm, and from which one can extract
global event shapes \emph{and} traditional jets. 
Furthermore, by utilizing more correlations from across the detector, 
it may be possible to treat pileup as a single entity that can be subtracted \emph{en massse}.
Not only could such an approach address the challenges of the HL-LHC, 
it might unlock new ways to study QCD.
In this paper, we demonstrate how the spectral power $H_l\ds$ can begin to 
fill this middle ground. Section \ref{sec:H_l} will define and introduce $H_l\ds$,
section \ref{sec:H_l-jets} will use $H_l\ds$ to define a jet definition, and
section \ref{sec:reco} will use $H_l\ds$ jets to reconstruct 
a few important final states.


\section{Spectral power $H_l\ds$}\label{sec:H_l}

To define the spectral power of collider final states,
we solicit the aid of a toy model. Imagine a collider where:
\begin{enumerate}
	\setlength\itemsep{0.em}	
	\item The lab frame is the center-of-momentum (CM) frame.
	\item All final-state particles are massless (i.e. $E=|\vec{p}\,|$).\label{masslessness}
	\item All interactions (including decays) occur within a nanoscopic region, 
		so that all final-state particles appear to converge at a single spacetime point $x^\mu_0$.
\end{enumerate}
Since particle radial position $r=c\,t$, the event can be fully described by
projecting it onto the unit sphere, with the momentum of individual particles
expressed in terms of a real, continuous, and non-negative energy density $\rho(\hat{r})$
\begin{equation}
	\vec{p}\,(\hat{r}) = \rho(\hat{r})\,\hat{r}\,,
\end{equation}
so that 
\begin{equation}\label{eq:Etot}
	E_\text{tot}\ds = \frac{1}{4\pi}\int \rho(\hat{r})\diff{\Omega}\,.
\end{equation}
This naturally suggests characterizing an event's shape and kinematics via the 
orthonormal spherical harmonics $Y_l^m$, most conveniently via 
the rotationally invariant power spectrum
\begin{equation}\label{eq:H_l-cont}
	H_l\ds = \frac{1}{\Etot^2}\frac{4\pi}{(2l+1)}\sum_{m=-l}^{+l}
	\left|\int Y_l^m(\hat{r})\rho(\hat{r})\diff{\Omega}\right|^2\,,
\end{equation}
which quantifies the magnitude of each multipole moment in the $Y_l^m$ decomposition.\footnote
{\nobreak
	$H_0\ds = 1$ by definition.
	If the lab frame is the CM frame, conservation of momentum requires $H_1\ds = 0$.
}
For example, one would na\"ively expect a 2-jet event to have 
most of its power in the dipole moment~$H_2\ds$.

The three properties of our toy model roughly approximate a lepton collider.
And since any experiment corresponds to a finite sampling of $\rho(\hat{r})$,
we must use a discrete energy density for a set of $N$ detected particles, 
each with energy fraction $f_i\ds \equiv E_i\ds / \Etot\ds$ and unit direction~$\hat{p}_i\ds$
\begin{equation}\label{eq:discrete-rho}
	\rho(\hat{r}) =\Etot\ds \sum f_i\ds\delta^3(\hat{r} - \hat{p}_i\ds)\,.
\end{equation}
The discrete density produces the power spectrum introduced by Fox and Wolfram~\cite{Fox:1978vu}
\begin{equation}\label{eq:H_l}
	 H_l\ds = \sum_{ij} f_i\ds f_j\ds P_l\ds(\hat{p}_i\ds \cdot \hat{p}_j\ds)
	  = \bra{f} P_l\ds\big(\,\ket{\hat{p}}\cdot\bra{\hat{p}}\,\big)\ket{f}\,,
\end{equation}
which uses Legendre polynomials $P_l\ds(x)$ to find 
two-particle angular/energy correlations.

The weight $w_{ij}\ds=f_i\ds f_j\ds$ of each 
angular contribution $P_l\ds(\hat{p}_i\ds \cdot \hat{p}_j\ds)$ 
makes $H_l\ds$ relatively insensitive to the addition of 
a handful of soft particles (which have $f\ll1$ by definition).
Similarly, a particle splitting to two nearly parallel particles ($a\to b\,c$) 
alters $H_l$ minimally because ${f_b\ds + f_c\ds = f_a\ds}$, 
so that the sum over $a$'s contributions before the splitting
\begin{equation}
	H_{l,a}\ds = f_a\ds \sum_{j}\ds f_j\ds P_l\ds(\hat{p}_a\ds \cdot \hat{p}_j)
\end{equation}
is only perturbed in the $P_l\ds$ terms after the splitting 
(where $\hat{p}_{b/c}\ds = \hat{p}_a\ds + \vec{\delta}_{b/c}$ for some $|\vec{\delta}_{b/c}|\ll1$)
\begin{align}
	H_{l,bc}\ds
	& = f_b\ds \sum_{j}\ds f_j\ds P_l\ds((\hat{p}_a\ds + \vec{\delta}_b\ds)\cdot \hat{p}_j)\nonumber \\
	& + f_c\ds \sum_{j}\ds f_j\ds P_l\ds((\hat{p}_a\ds + \vec{\delta}_c\ds)\cdot \hat{p}_j)\,.
\end{align}
Only when $l$ becomes large --- and $P_l\ds(x+\delta x)$ highly oscillatory --- 
can a small $\delta x$ give rise to significant changes in~$H_l\ds$.
This allows us to deduce that, for reasonably coarse event shapes (low to moderate~$l$),
the power spectrum is infrared and collinear (IRC) safe.

At small angles the discrete power spectrum loses IRC safety,
while also becoming dominated by sampling noise. 
In this large-$l$ regime, $H_l\ds$ can be meaningfully decomposed into 
two types of correlations; a constant term from particle self correlations
(via ${P_l\ds(\hat{p}_i\ds \cdot \hat{p}_i\ds) = 1}$) and
an ${l\text{-dependent}}$, inter-particle correlation term
\begin{equation}
	H_l\ds = \underset{\text{self}}{\underbrace{\braket{f}{f}}} +
		\underset{\text{inter-particle}}{\underbrace{
		\bra{f}\left\lbrack P_l\ds\big(\,\ket{\hat{p}}\cdot\bra{\hat{p}}\,\big) - \ident\right\rbrack\ket{f}}}\,.
\end{equation}
The latter tends to destructively interfere
for large $l$ (small angles), so that $H_l\ds$ 
asymptotically flattens to a finite floor that persist as $l\to\infty$
\begin{equation}\label{eq:H_l-asymp}
	H_l\ds\to\braket{f}{f}\lbrack1\pm(\text{small fluctuations})\rbrack.
\end{equation}
For ``well-behaved'' event topologies 
(those where the energy fraction of individual particles follow a 
probability density $g(f)$ with finite variance),\footnote
{\nobreak 
	Throughout, $g(x)$ is a smooth, normalized probability density function for variable $x$.
}
\begin{equation}\label{eq:ff-prop}
	\braket{f}{f} \propto N^{-1}\,.
\end{equation}
Thus, the finite power at large $l$ in Eq.~\ref{eq:H_l-asymp} is 
inversely proportional to particle multiplicity.

This constant asymptotic power $\braket{f}{f}$,
which guarantees that ${\sum H_l\ds = \infty}$, 
can be interpreted as \emph{white noise}.
Equation~\ref{eq:discrete-rho}'s discrete energy distribution $\rho(\hat{r})$ 
is built from weighted delta functions, which the spherical harmonics
approximate by adding ever-higher orders of $Y_l^m$.
Reproducing the infinitesimal width of each $\delta$ requires 
a significant amount of spectral power as $l\to\infty$.
Furthermore, the discrete nature of the sample creates unreliable (IRC unsafe)
small-angle correlations between the delta functions,
which leads to the ``small fluctuations'' in the asymptotic $H_l\ds$.
Larger particle multiplicity lowers the white noise floor because
the individual delta functions in $\rho(\hat{r})$ have smaller weights, 
requiring less power to reconstruct. 
Furthermore, the finer sampling of the underlying energy distribution $\rho(\hat{r})$
pushes unreliable correlations to larger $l$. 
Hence, observed particle multiplicity limits the angular resolution of the spectral power.

\begin{figure}[htb]
\subfloat[\label{fig:2-jet}]{\includegraphics[scale=1]{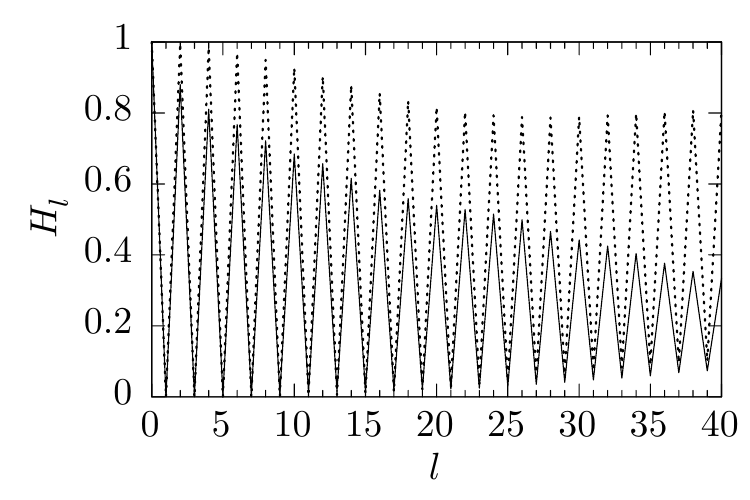}}
\subfloat[\label{fig:3-jet}]{\includegraphics[scale=1]{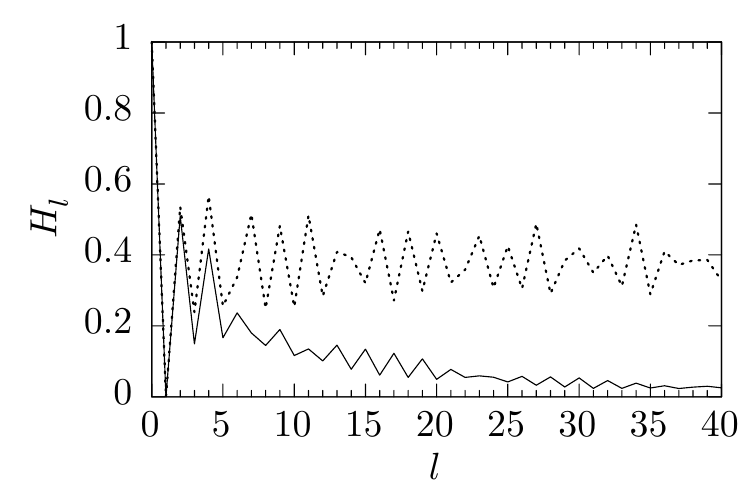}}
\caption{The spectral power $H_l\ds$ for two $e^+e^-\to q\bar{q}g$ events,
(a) one which is very 2-jet-like and (b) one which is distinctly 3-jet-like. 
$H_l\ds$ is shown for both the 3-parton matrix element (dotted) and the 
$\bigO{100}$-particle final state after showering and hadronization (solid). 
The power spectrum is only defined for integer $l$, but connecting lines are added to aid the eye.}
\label{fig:2/3-jet}
\end{figure}

We can see the effect of multiplicity in Fig.~\ref{fig:3-jet}, 
which shows the spectral power for an $e^+e^-\to q\bar{q}g$ event.
The $H_l\ds$ for its 3-parton matrix element (dotted) is highly oscillatory,
and because there are only three particles it maintains a 
large average power as $l\to\infty$.
The much higher multiplicity of its showered final state gradually decreases 
the spectral power until it reaches its asymptotic floor.
Note that there are no easily identifiable patterns in the spectral power
for either the matrix element or showered final state,
and that the similarities between them quickly diminish after the first few orders of~$l$.

In Fig.~\ref{fig:3-jet}, the gluon in the matrix element carries $f=16\%$ of the event's
energy, making it a noticeably 3-jet-like event. In Fig.~\ref{fig:2-jet}, 
the gluon carries a much lower fraction $f=6\%$, making it a much more 2-jet-like event. 
Comparing the showered $H_l\ds$ (solid) between the two events, 
the 3-jet-like event clearly has a larger value of $H_3\ds$ than the 2-jet-like event.
However, note that the 3-jet-like event is still dominated by $H_2\ds$ and $H_4\ds$.
(and the 2-jet-like event a significant $H_2\ds$).
The spectral power of QCD jets does not behave as one might na\"ively assume.

\section{$H_l\ds$ jets}\label{sec:H_l-jets}

Spectral power is not novel. It is used to quantify the 
angular scale of fluctuations in the cosmic microwave background (CMB).
Using relatively simple models, the broad peaks in the CMB power spectrum 
have been used quite successfully to infer 
important cosmological properties of the universe.
One can imagine using $H_l\ds$ in a similar fashion, 
to map out high-frequency information such as jet substructure.
However, there is an immediate problem with this approach;
unlike the CMB, collider events are \emph{not} a mostly isotropic sphere.
They consist of discrete, collimated objects and vast swaths of inactive surface.
As such, high-frequency correlations inside a jet are always localized, 
so any broad shapes at large $l$ from jet substructure are conflated with a 
spatial filter surrounding each jet.
The resulting $H_l\ds$ is highly oscillatory, as we saw in Fig.~\ref{fig:2/3-jet}
for simple multijet events.

When Fox and Wolfram first applied spectral power to collider events, 
they faced the additional complication that event multiplicity was low, 
shrouding high-frequency information in sampling noise.
Being restricted to low-$l$ information, they calculated
the probability densities $g(H_l\ds)$ for $l=2\ds$ and 3 
expected from three different differential cross sections
($e^+e^-\to q\bar{q}$, $e^+e^-\to q\bar{q}g$, and $e^+e^- \to X_\text{heavy} \to ggg$), 
in order to distinguish these final states using $H_2\ds$ and $H_3\ds$~\cite{Fox:1978vu}.
This is certainly useful, but it neglects the 
significant changes to $H_l\ds$ (even at low $l$) that can arise from
parton showering, which significantly alters an event's shape.
Furthermore, limiting ones attention to the first few orders 
throws out much of the benefit of using spectral power in the first place 
--- utilizing all the information in the detector.

Since we now inhabit a high multiplicity environment, 
our initial foray into the spectral power of collider physics takes a different route.
We calculate the spectral power $H_l^\text{obs}$ for a set of $N$ observed particles,
then ``reverse engineer'' a set of $n$ massless partons ($n\ll N$) whose 
spectral power $H_l^\text{reco}$ reproduces $H_l^\text{obs}$ from $l=0$ to $l_\text{max}\ds$.
Starting with $l_\text{max}\ds = 2$, we can 
increment $l_\text{max}\ds\to l_\text{max}\ds + 1$ after each fit
to gently add high-$l$ information (fine structure) to the existing reconstruction.
The result is an $n$-jet reconstruction which focuses on the coarse event structure, 
but does not assign specific particles to specific jets, 
and simultaneously uses every 2-particle correlation at every stage of the reconstruction.

To ``reverse engineer'' $H_l^\text{obs}$, we minimize the square residuals
\begin{equation}\label{eq:sum-r}
	S=\sum _{l=0}^{l_\text{max}\ds}r_l^2,
	\qquad\text{with}\quad 
		r_l\ds \equiv H_l^\text{reco} - H_l^\text{obs}\,.
\end{equation}
We must then choose a set of parameters $\vec{b}$ to characterize the $n$ jets.
Retaining our toy model, the number of parameters is
\begin{equation}\label{eq:num-param}
	\text{len}\left(\vec{b}\right) = 
	\begin{cases}
		0 & n = 2\\
		2 + 3(n-3) & n > 2 \\
	\end{cases}
	\,.
\end{equation}
$H_l\ds$ is invariant to the absolute event orientation,
which fixes the first jet to an arbitrarily chosen axis.
And since we are observing in the event's CM frame, 
the final jet is fixed by momentum conservation.
This leaves zero parameters for $n=2$.
For $n=3$, full freedom is given to the second jet's 3-momentum, 
from which the azimuthal symmetry of $H_l\ds$ removes one degree of freedom.
Each additional jet adds a full 3 d.o.f.

The simplest choice for parameters is to use the actual components of the 
$(n-2)$ free jets' 3-momenta~$\vec{p}_i\ds$;
\begin{equation}
	\vec{b} = \lbrace p_2^{(2)}, p_2^{(3)}, \dots, p_{N-1}^{(1)}, p_{N-1}^{(2)}, p_{N-1}^{(3)}\rbrace
\end{equation}
(where we arbitrarily choose $\vec{p}_1\ds=\hat{z}$ and constrain $\vec{p}_2\ds$ to the $yz$ plane).
With this parameterization, it is relatively straightforward to infer that the Jacobian of the fit
\begin{equation}
	J_{lk}\ds = -\frac{\partial r_l\ds}{\partial b_k\ds}
\end{equation}
is highly non-linear. A non-linear least squared (NLLS) minimization algorithm
can only guarantee a local minimum, which demands a clever choice of $\vec{b}_0\ds$.
A good source of $\vec{b}_0\ds$ is a traditional, 
sequential jet reconstruction.
One then solves for the rotation matrix which simultaneously takes 
the leading jet to~$\hat{z}$ and the sub-leading jet to the $yz$ plane,
which allows the restoration of the absolute orientation after the fit.


\section{Final state reconstruction}\label{sec:reco}

To test the efficacy of $H_l\ds$ jets we use a toy detector
with polar angle $\theta$ and azimuthal angle $\phi$,
loosely based on experiments at the LHC.
The detector perfectly sees all charged particles with $p_T\ds > 150\;\unit{MeV}$ out to a pseudorapidity $|\eta|<2.5$,
but can only resolve neutral particles via a calorimeter which extends to $|\eta|<5$.
The calorimeter is segmented into square towers of angular dimension
$\Delta\theta\times\Delta\phi$ arranged into belts of constant $\theta$
which azimuthally wrap the collision point. 
Each belt uses $\Delta\theta = 0.1$ and 
$\Delta\phi \approx \frac{\Delta\theta}{\sin(\theta)}$, 
so that each tower covers an approximately equivalent 
differential solid angle $\diff{\Omega} = (\Delta \theta)^2$.
The calorimeter accumulates neutral particle energy (track-subtracted)
with perfect accuracy, returning a massless momentum $p^\mu$ at the center of each struck tower. 
While this detector scheme is admittedly crude,
its purpose is to take a truth-level reconstruction
and filter the small-angle correlations in the neutral sector,
reducing our sensitivity to fundamentally unattainable information.

We simulate $e^+e^-$ events at $\sqrt{s}=400\;\unit{GeV}$,
using \textsc{MadGraph~5} to generate the matrix element~\cite{Alwall:2014hca}
and \textsc{Pythia~8} to perform jet showering and hadronization~\cite{Sjostrand:2006za, Sjostrand:2007gs}.
The seeds for $H_l\ds$ jets are generated via the anti-$k_t\ds$ algorithm
implemented in \texttt{FastJet~3}~\cite{Cacciari:2011ma}, with clustering radius $R=0.4$. 
The $H_l\ds$ jet fit starts with $l_\text{max} = 2$ 
and keeps incrementing $l_\text{max}$ so long as the sum of residuals remains small
($S < (\max(H_l\ds)/20)^2$ for $l>0$; the maximum allowable $S$ is 
scaled to the strongest power in the event).

We first fit both of the $e^+e^-\to q\bar{q}g$ events 
shown in Fig.~\ref{fig:2/3-jet} using an $n=3$ jet fit.
In the CM frame, the kinematics of a 3-jet system (minus the absolute orientation) 
are fully described by the energy fractions of the two leading jets (2 d.o.f.).
Hence, the figure of merit is the fitted energy fractions $f$ versus those in the matrix element. 
However, since $H_l\ds$ cannot distinguish a quark from a gluon 
(at least not without utilizing spin correlations which may or may not exist at large~$l$), 
we simply choose the two largest $f$ in each event.

\begin{figure}[h]
\subfloat[\label{fig:2-jet-fit}]{\includegraphics[scale=1]{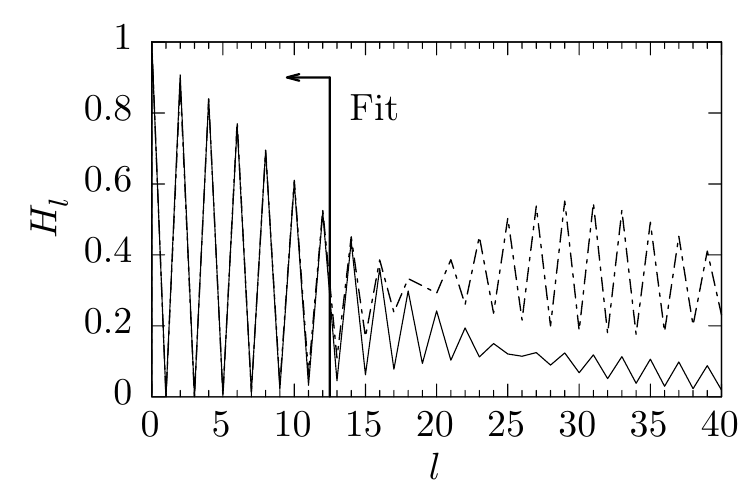}}
\subfloat[\label{fig:3-jet-fit}]{\includegraphics[scale=1]{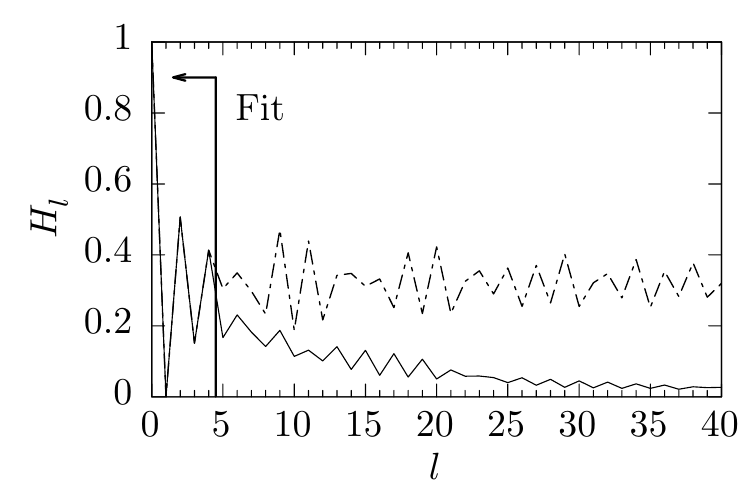}}
\caption{The spectral power $H_l\ds$ for the two events depicted in Fig.~\ref{fig:2/3-jet}
given (solid) the $\bigO{100}$ tracks/towers in the detector and (dot-dashed) the 3-jet fit.
The arrow depicts $l_\text{max}\ds$.}
\label{fig:2/3-jet-fit}
\end{figure}

In Fig.~\ref{fig:2/3-jet-fit}, the fit can only accommodate $l_\text{max}=3$
for the 3-jet-like event, but it is nonetheless able to 
reconstruct the leading jet's $f$ within $+6\%$ and 
the sub-leading jet within $-6\%$ (with an accurate third jet).
The fit for the 2-jet-like event is able to use
many more data points with a good residual $(l_\text{max}=12)$,
and while the leading jet's $f$ is reconstructed within $0.3\%$, 
the minuscule third jet is replaced by a sub-leading jet which
splits into two collinear subjets with nearly equal~$f$.

Our results lead directly to an important point.
Equations~\ref{eq:H_l-asymp}~and~\ref{eq:ff-prop} impose a 
strict limitation on $H_l\ds$ jets. A low-multiplicity $n$-jet system will reproduce an 
$H_l^\text{reco}$ with a large asymptotic value of 
$H_l^\text{reco}\sim\braket{f}{f}\propto n^{-1}$, 
whereas the spectral power of the high multiplicity observation
more quickly attenuates towards a significantly lower~$H_l^\text{obs}\sim N^{-1}$.
This mismatch causes the reconstructed spectral power to hover above the observation,
creating a forcing during the fit that drives the fit parameters towards
a smaller value of~$\braket{f}{f}$.
It is relatively easy to show that the smallest possible 
${\braket{f}{f}_\text{min}\ds = n^{-1}}$ corresponds to the
rather trivial configuration where all $n$ jets have the same energy fraction.
Thus, to mimic the multiplicity attenuation in $H_l^\text{obs}$ at large $l$,
the fit will tend to ``equilibrate'' the energy fraction of the $H_l\ds$ jets.
This is exactly what occurs in the 2-jet-like system;
with a 3-jet fit, a puny third jet is better reconstructed as a collinear splitting, 
since this does a better job at reducing $\braket{f}{f}$ to match the observation, 
permitting $l_\text{max}$ to increment to large value.
For the 3-jet-like system, the large value of $H_3\ds$ 
demands a independent third jet, giving an $H_l^\text{reco}$
that does not attenuate fast enough to match observations beyond $l_\text{max}=4$.

\begin{figure}[htb]
\subfloat[\label{fig:tt-overfit}]{\includegraphics[scale=1]{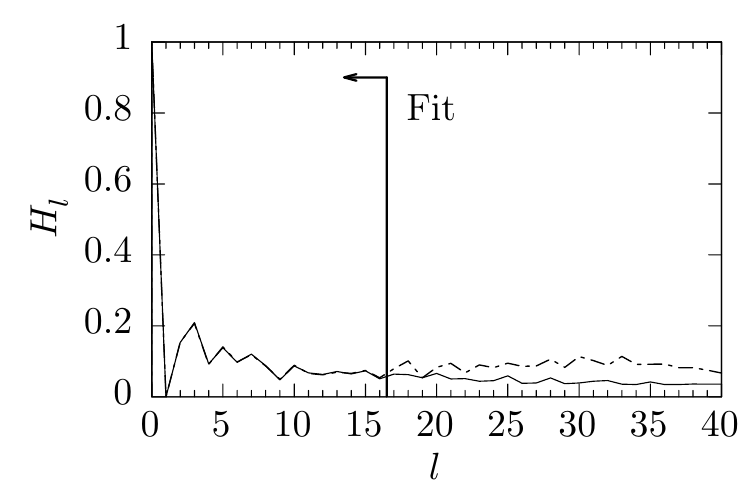}}
\subfloat[\label{fig:3-jet-overfit}]{\includegraphics[scale=1]{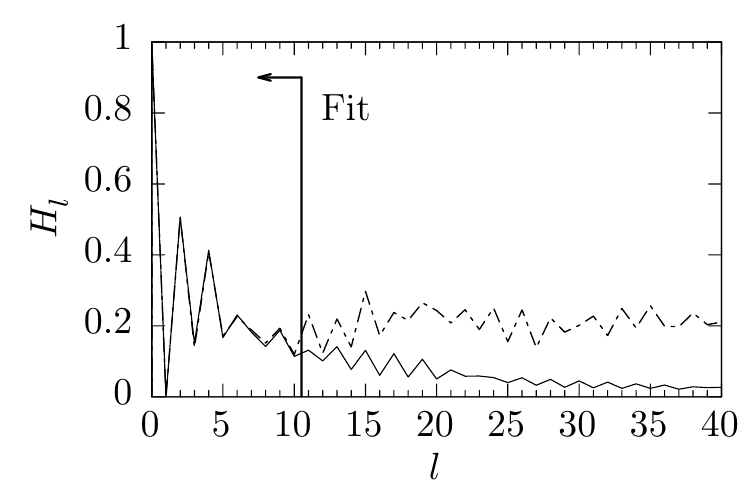}}
\caption{The spectral power $H_l\ds$ for (a) a $t\bar{t}$ event (6-jet) and 
(b) a 3-jet-like event. Both events use a 7-jet fit (dot-dashed).}
\label{fig:overfit}
\end{figure}

While increasing $n$ ostensibly solves this problem 
--- more jets in the fit create more multiplicity attenuation ---
it turns out that new jet parameters accumulate
faster than the increase in $l_\text{max}$ which they allow.
This is evident in Fig.~\ref{fig:overfit},
which shows an $n=7$ jet fit for the same 3-jet-like event (Fig.~\ref{fig:3-jet-overfit})
as well as an $e^+e^-\to t\bar{t}\to6\,\text{jet}$ event (Fig.~\ref{fig:tt-overfit}).
In both cases, the large $n$ permits more observations in the fit,
yet not enough to effectively constrain the 14 parameters of a 7 jet system (Eq.~\ref{eq:num-param})
--- they are both highly underfit.
This becomes even clearer when one notices that 
the reconstructed spectral power abruptly increases to 
a much larger power immediately after the last fitted point,
indicating a pathological solution where the energy fraction of the jets
has been equilibrated due to an artificial constraint (keeping $n\ll N$).

This problem motivates the development of a jet shape parameter to
distribute the energy of the $n$ jets in the $H_l\ds$ reconstruction, making them extensive objects.
A jet shape function morphs the $\delta$-functions of the energy density $\rho(\hat{r})$
into a collection of more diffuse 
distributions. It is easier to reproduce such shapes with lower-$l$ spherical harmonics,
so less power is needed at large $l$. 
Because extensive jets have less ``white noise,'' $H_l^\text{reco}$ is
given the multiplicity attenuation it needs.
In addition to improving the fit of $H_l\ds$ jets,
jet shapes offer an avenue for empirically measuring the shape of real jets
using $N^2$ correlations from across the detector,
then comparing those measurements to the predictions of QCD.

\subsection{Pileup}

We add random ``pileup'' to the detector using a simplistic model; 
pileup is isotropic, it only appears in towers (e.g.\ charged pileup is removable), 
and its energy follows an exponential distribution 
$g(E) = \lambda \exp(-\lambda E)$ with $\lambda = 10^3/\sqrt{s}$
(so that $\langle E\rangle = 0.1\%$ of the collider energy).
The isotropic nature of this model ensures that pileup's 
dominant power should be $H_0\ds$, 
with a featureless floor of white noise for $l>0$.
Thus, pileup merely attenuates the energy fraction of the signal, 
globally scaling its power spectrum by $(1-f_\text{pu})^2$, 
so that pileup can be added to $H_l$ jets via a single free parameter,
the pileup energy fraction $f_\text{pu}$.

\begin{figure}[htb]
\subfloat[\label{fig:2-jet-fit}]{\includegraphics[scale=1]{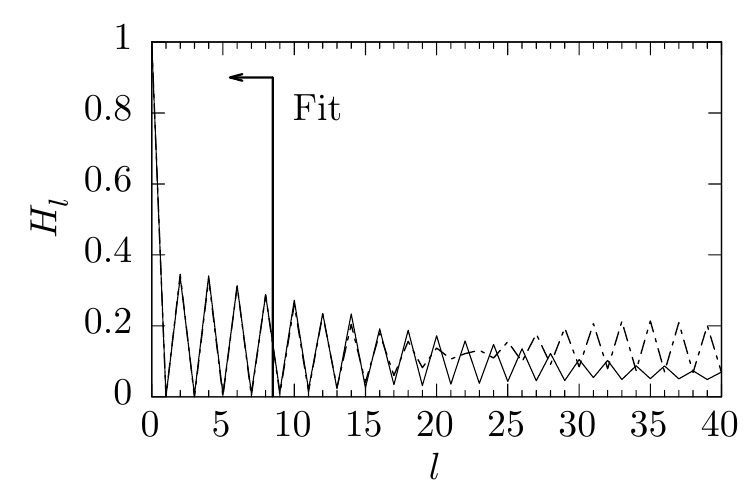}}
\subfloat[\label{fig:3-jet-fit}]{\includegraphics[scale=1]{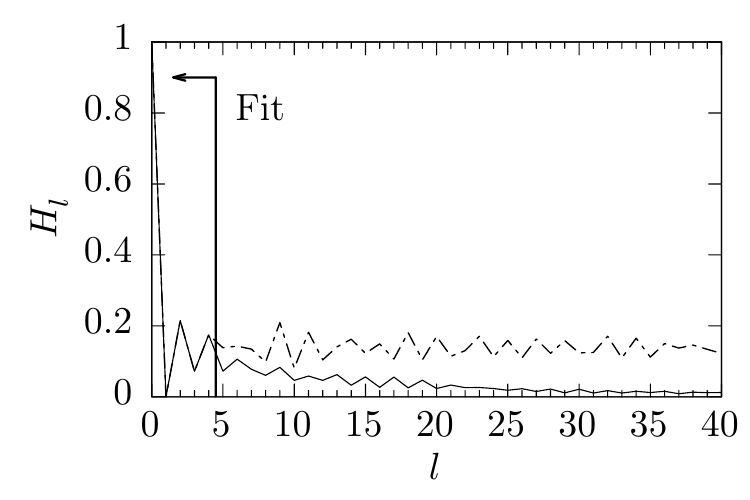}}
\caption{The spectral power $H_l\ds$ for the two events depicted in Fig.~\ref{fig:2/3-jet-fit}, 
with (solid) isotropic pileup $(S/N = 2/1)$ in the detector and (dot-dashed) the 3-jet fit.
The arrow depicts $l_\text{max}\ds$.}
\label{fig:2/3-jet-pu-fit}
\end{figure}

In Fig.\ref{fig:2/3-jet-pu-fit} we add a very significant amount of pileup $(S/N=2/1)$
to the events previously depicted in Fig.~\ref{fig:2/3-jet-fit},
and are nonetheless able to reconstruct both systems with
equivalent kinematics to the no-pileup fit $(\pm1\%)$.
This implies that $H_l\ds$ jets can accommodate an arbitrary pileup shape
provided that the spectral power of pileup itself can be measured in experiment
(e.g.\ by averaging over many ``min bias'' events and assuming the 
shape of pileup is consistent for every event).
This measured $H_l^\text{pu}$ can then be scaled by $f_\text{pu}$ and 
subtracted from $H_l^\text{obs}$ during the fit. 
This scheme allows the significant angular-energy correlations of 
pileup itself to smooth out its perturbations to the
spectral power of the signal, permitting its subtraction as a whole entity.

\section{Conclusion}

We revisit the spectral power $H_l\ds$ and its applications to studying 
multiparticle final states at a particle collider.
We define the $H_l$ jet definition, and our preliminary results
demonstrate a promising ability to correctly identify jet kinematics, 
even in the face of significant pileup $(S/N = 2/1)$.
However, an $n$-jet reconstruction is too ``needle-like,'' 
and future work will investigate shape parameters to make jets 
extensive objects distributed in space. Furthermore, 
the ultimate goal of applying $H_l\ds$ jets at the LHC 
will require addressing the unknown longitudinal boost of the CM frame in a proton collider.


\bibliography{dpf2017_proceeding}

\begin{thebibliography}{5}%
\makeatletter
\providecommand \@ifxundefined [1]{%
 \@ifx{#1\undefined}
}%
\providecommand \@ifnum [1]{%
 \ifnum #1\expandafter \@firstoftwo
 \else \expandafter \@secondoftwo
 \fi
}%
\providecommand \@ifx [1]{%
 \ifx #1\expandafter \@firstoftwo
 \else \expandafter \@secondoftwo
 \fi
}%
\providecommand \natexlab [1]{#1}%
\providecommand \enquote  [1]{``#1''}%
\providecommand \bibnamefont  [1]{#1}%
\providecommand \bibfnamefont [1]{#1}%
\providecommand \citenamefont [1]{#1}%
\providecommand \href@noop [0]{\@secondoftwo}%
\providecommand \href [0]{\begingroup \@sanitize@url \@href}%
\providecommand \@href[1]{\@@startlink{#1}\@@href}%
\providecommand \@@href[1]{\endgroup#1\@@endlink}%
\providecommand \@sanitize@url [0]{\catcode `\\12\catcode `\$12\catcode
  `\&12\catcode `\#12\catcode `\^12\catcode `\_12\catcode `\%12\relax}%
\providecommand \@@startlink[1]{}%
\providecommand \@@endlink[0]{}%
\providecommand \url  [0]{\begingroup\@sanitize@url \@url }%
\providecommand \@url [1]{\endgroup\@href {#1}{\urlprefix }}%
\providecommand \urlprefix  [0]{URL }%
\providecommand \Eprint [0]{\href }%
\providecommand \doibase [0]{http://dx.doi.org/}%
\providecommand \selectlanguage [0]{\@gobble}%
\providecommand \bibinfo  [0]{\@secondoftwo}%
\providecommand \bibfield  [0]{\@secondoftwo}%
\providecommand \translation [1]{[#1]}%
\providecommand \BibitemOpen [0]{}%
\providecommand \bibitemStop [0]{}%
\providecommand \bibitemNoStop [0]{.\EOS\space}%
\providecommand \EOS [0]{\spacefactor3000\relax}%
\providecommand \BibitemShut  [1]{\csname bibitem#1\endcsname}%
\let\auto@bib@innerbib\@empty
\bibitem [{\citenamefont {Fox}\ and\ \citenamefont
  {Wolfram}(1978)}]{Fox:1978vu}%
  \BibitemOpen
  \bibfield  {author} {\bibinfo {author} {\bibfnamefont {G.~C.}\ \bibnamefont
  {Fox}}\ and\ \bibinfo {author} {\bibfnamefont {S.}~\bibnamefont {Wolfram}},\
  }\href {\doibase 10.1103/PhysRevLett.41.1581} {\bibfield  {journal} {\bibinfo
   {journal} {Phys. Rev. Lett.}\ }\textbf {\bibinfo {volume} {41}},\ \bibinfo
  {pages} {1581} (\bibinfo {year} {1978})}\BibitemShut {NoStop}%
\bibitem [{\citenamefont {Alwall}\ \emph {et~al.}(2014)\citenamefont {Alwall},
  \citenamefont {Frederix}, \citenamefont {Frixione}, \citenamefont {Hirschi},
  \citenamefont {Maltoni}, \citenamefont {Mattelaer}, \citenamefont {Shao},
  \citenamefont {Stelzer}, \citenamefont {Torrielli},\ and\ \citenamefont
  {Zaro}}]{Alwall:2014hca}%
  \BibitemOpen
  \bibfield  {author} {\bibinfo {author} {\bibfnamefont {J.}~\bibnamefont
  {Alwall}}, \bibinfo {author} {\bibfnamefont {R.}~\bibnamefont {Frederix}},
  \bibinfo {author} {\bibfnamefont {S.}~\bibnamefont {Frixione}}, \bibinfo
  {author} {\bibfnamefont {V.}~\bibnamefont {Hirschi}}, \bibinfo {author}
  {\bibfnamefont {F.}~\bibnamefont {Maltoni}}, \bibinfo {author} {\bibfnamefont
  {O.}~\bibnamefont {Mattelaer}}, \bibinfo {author} {\bibfnamefont {H.~S.}\
  \bibnamefont {Shao}}, \bibinfo {author} {\bibfnamefont {T.}~\bibnamefont
  {Stelzer}}, \bibinfo {author} {\bibfnamefont {P.}~\bibnamefont {Torrielli}},
  \ and\ \bibinfo {author} {\bibfnamefont {M.}~\bibnamefont {Zaro}},\ }\href
  {\doibase 10.1007/JHEP07(2014)079} {\bibfield  {journal} {\bibinfo  {journal}
  {JHEP}\ }\textbf {\bibinfo {volume} {07}},\ \bibinfo {pages} {079} (\bibinfo
  {year} {2014})},\ \Eprint {http://arxiv.org/abs/1405.0301} {arXiv:1405.0301
  [hep-ph]} \BibitemShut {NoStop}%
\bibitem [{\citenamefont {Sjostrand}\ \emph {et~al.}(2006)\citenamefont
  {Sjostrand}, \citenamefont {Mrenna},\ and\ \citenamefont
  {Skands}}]{Sjostrand:2006za}%
  \BibitemOpen
  \bibfield  {author} {\bibinfo {author} {\bibfnamefont {T.}~\bibnamefont
  {Sjostrand}}, \bibinfo {author} {\bibfnamefont {S.}~\bibnamefont {Mrenna}}, \
  and\ \bibinfo {author} {\bibfnamefont {P.~Z.}\ \bibnamefont {Skands}},\
  }\href {\doibase 10.1088/1126-6708/2006/05/026} {\bibfield  {journal}
  {\bibinfo  {journal} {JHEP}\ }\textbf {\bibinfo {volume} {05}},\ \bibinfo
  {pages} {026} (\bibinfo {year} {2006})},\ \Eprint
  {http://arxiv.org/abs/hep-ph/0603175} {arXiv:hep-ph/0603175 [hep-ph]}
  \BibitemShut {NoStop}%
\bibitem [{\citenamefont {Sjostrand}\ \emph {et~al.}(2008)\citenamefont
  {Sjostrand}, \citenamefont {Mrenna},\ and\ \citenamefont
  {Skands}}]{Sjostrand:2007gs}%
  \BibitemOpen
  \bibfield  {author} {\bibinfo {author} {\bibfnamefont {T.}~\bibnamefont
  {Sjostrand}}, \bibinfo {author} {\bibfnamefont {S.}~\bibnamefont {Mrenna}}, \
  and\ \bibinfo {author} {\bibfnamefont {P.~Z.}\ \bibnamefont {Skands}},\
  }\href {\doibase 10.1016/j.cpc.2008.01.036} {\bibfield  {journal} {\bibinfo
  {journal} {Comput. Phys. Commun.}\ }\textbf {\bibinfo {volume} {178}},\
  \bibinfo {pages} {852} (\bibinfo {year} {2008})},\ \Eprint
  {http://arxiv.org/abs/0710.3820} {arXiv:0710.3820 [hep-ph]} \BibitemShut
  {NoStop}%
\bibitem [{\citenamefont {Cacciari}\ \emph {et~al.}(2012)\citenamefont
  {Cacciari}, \citenamefont {Salam},\ and\ \citenamefont
  {Soyez}}]{Cacciari:2011ma}%
  \BibitemOpen
  \bibfield  {author} {\bibinfo {author} {\bibfnamefont {M.}~\bibnamefont
  {Cacciari}}, \bibinfo {author} {\bibfnamefont {G.~P.}\ \bibnamefont {Salam}},
  \ and\ \bibinfo {author} {\bibfnamefont {G.}~\bibnamefont {Soyez}},\ }\href
  {\doibase 10.1140/epjc/s10052-012-1896-2} {\bibfield  {journal} {\bibinfo
  {journal} {Eur. Phys. J.}\ }\textbf {\bibinfo {volume} {C72}},\ \bibinfo
  {pages} {1896} (\bibinfo {year} {2012})},\ \Eprint
  {http://arxiv.org/abs/1111.6097} {arXiv:1111.6097 [hep-ph]} \BibitemShut
  {NoStop}%
\end{thebibliography}%

\end{document}